\documentstyle[epsf]{aipproc2}
\def\thefootnote{\fnsymbol{footnote}}
\def\bea {\begin{eqnarray}}
\def\eea {\end{eqnarray}}
\def\be {\begin{equation}}
\def\ee {\end{equation}}
\def\ben{\begin{enumerate}}
\def\een{\end{enumerate}}
\def\bi{\begin{itemize}}
\def\ei{\end{itemize}}
\def\ie{{\it i.e.}}

\def\etal{{\it et al.}}

\def\F{{\cal F}}

\def\prl {{\it Phys. Rev. Lett.\ }}
\def\plb {{\it Phys. Lett.} B}
\def\pr {{\it Phys. Rev.\ }}
\def\prc {{\it Phys. Rev.} C}
\def\prd {{\it Phys. Rev.} D}
\def\npa {{\it Nucl. Phys.} A}
\def\epjc{{\em Eur. Phys. J.} C}

\def\GV{G_{\mbox{\tiny V}}}
\def\GF{G_{\mbox{\tiny F}}}
\def\DRV{\Delta_{\mbox{\tiny R}}^{\mbox{\tiny V}}}

\def\mA{m_{\mbox{\tiny A}}}

\def\mZ{m_{\mbox{\tiny Z}}}

\def\MV{M_{\mbox{\tiny V}}}

\def\mids{\! \mid \! }
\def\hyphen{{\mbox{-}}}

\begin{document}

\title{Superallowed Fermi Beta Decay \\ and Coulomb
Mixing in Nuclei} 
\author{J.C. Hardy$^{a}$ and I. S. Towner$^{b}$ }
\address{$^{a}${Cyclotron Institute, Texas A \& M University, \\
College Station, TX 77843} \\
$^{b}${Physics department, Queen's University, \\
Kingston, Ontario K7L 3N6, Canada}}
\maketitle

\begin{abstract}
     Superallowed $0^+ \rightarrow 0^+$ nuclear
beta decay provides a direct measure of the weak
vector coupling constant, $\GV$.  We survey
current world data on the nine accurately
determined transitions of this type, which range
from the decay of $^{10}$C to that of $^{54}$Co,
and demonstrate that the results confirm
conservation of the weak vector current (CVC) but
differ at the 98\% confidence level from the
unitarity condition for the Cabibbo-Kobayashi-Maskawa 
(CKM) matrix.  We examine the
reliability of the small calculated corrections that
have been applied to the data, and conclude that there
are no evident defects although the Coulomb
correction, $\delta_C$, depends sensitively on nuclear
structure and thus needs to be constrained
independently.  The potential importance of a result in
disagreement with unitarity, clearly indicates the need
for further work to confirm or deny the discrepancy. 
We examine the options and recommend priorities for
new experiments  and improved calculations.  Some of
the required experiments depend upon the availability
of intense radioactive beams.  Others are possible with
existing facilities.

\end{abstract}

\renewcommand{\thefootnote}{\#\arabic{footnote}}
\setcounter{footnote}{0}

\section*{Introduction} \label{intro}

In probing the properties of the weak interaction, superallowed
$0^{+} \rightarrow 0^{+}$ nuclear beta decay has become a
singularly valuable tool, principally because of its relative
independence from the effects of nuclear structure, which are
notoriously difficult to account for with high accuracy.  Since
the measured $0^{+} \rightarrow 0^{+}$ transitions are
between $T=1$ analog states, nuclear-structure effects only
enter at the level of the differences between the parent and
daughter wave functions.  These differences, which are
caused by Coulomb and charge-dependent nuclear forces,
turn out to be very small, and introduce a correction of
order $1\%$ when the experimental $ft$-values are used to
extract a value for the effective weak vector coupling
constant, $\GV^{\prime}$.  Even a conservative estimate
of the uncertainties in this correction indicate that
structure-dependent uncertainties should not afflict the
experimental determination of $\GV^{\prime}$ above the
level of approximately $\pm 0.1\%$.  As a result,
considerable effort has gone into making $ft$-value
measurements that achieve this level of experimental
precision or better.

In this paper, we begin by summarizing the current
status of
world data on superallowed Fermi beta decays and
demonstrating the extent to which these data test the
Conserved Vector Current (CVC) hypothesis and the
unitarity of the Cabibbo-Kobayashi-Maskawa (CKM)
matrix.  In fact, we show that the success of the
measurements has been such that experiment has
outstripped theory.  The theoretical uncertainties in
calculated corrections -- including the 
nuclear-structure-dependent charge-corrections -- are
now the limiting
factor when superallowed beta decay is used to test the
validity of the Standard Model.  On the one hand, this
can be viewed as a serious limitation to the continued
usefulness of such measurements in future or, at least, as
an indication that any improvements in their precision will have
to be used to probe charge-dependence in nuclear structure
rather than the fundamental properties of weak
interactions.  On the other hand, it can also be taken as a
challenge to nuclear-structure theorists and experimenters
alike to establish the effects of charge-dependence by
independent means and thus to improve our ability overall to
calculate them precisely.

Following our survey of world data, we outline the current
approach to calculating charge-dependent effects, describe
measurements that have been used to test these
calculations independently, and consider future prospects
for improved results.  In particular, we shall consider the
prospects for enlarging the sample of well-measured
$0^{+}, T=1$ superallowed emitters, and examine what
needs to be done before such measurements can usefully
contribute to our fundamental understanding of the weak
interaction.

\section*{Current Status of World Data} \label{wdata}

Superallowed Fermi $0^+ \rightarrow 0^+$ nuclear beta
decays
\cite{Ha90,TH95}
provide both the best test of the CVC
hypothesis in weak interactions and, together with the
muon
lifetime, the most accurate value for the up-down
quark-mixing
matrix element of the CKM matrix,
$V_{ud}$.  Because the axial current cannot contribute in
lowest order to transitions between spin-0 states, the
experimental $ft$-value is related directly to the vector
coupling constant.  Specifically, for an
isospin-1 multiplet,

\be
ft (1 + \delta_R ) = \frac{K}{{\GV^{\prime}}^2 \langle \MV \rangle^2},
\label{ft}
\ee

\noindent with

\bea
\GV^{\prime} & = & \GV (1 + \DRV )^{1/2},
\nonumber \\
\langle \MV \rangle^2 & = & 2(1 - \delta_C),
\nonumber \\
K/(\hbar c)^6 & = & 2 \pi^3 \hbar \ln 2 / (m_e c^2)^5 =
( 8120.271 \pm 0.012) \times 10^{-10} GeV^{-4} s,
\label{MK}
\eea

\noindent where $f$ is the statistical rate function, $t$ is the
partial
half-life for the transition, $\langle \MV \rangle$ is the Fermi
matrix element and $\GV$ is the primitive vector
coupling constant.  The physical constants used to
evaluate $K$ were taken from the most recent Particle
Data Group publication \cite{PDG98}.  These
equations also include three calculated correction terms
-- all of order $1\%$.  We write $\delta_R$ as the
nucleus-dependent part of the radiative
correction, $\DRV $ as the nucleus-independent part of
the radiative correction, and $\delta_C$ as the isospin
symmetry-breaking correction.  A general description of
these three correction terms and the methods used in
their calculation has appeared elsewhere
\cite{Ha90,TH95,WEIN98}.  In the present context, it is
sufficient to note that nuclear structure plays
a small role in the determination of $\delta_R$, but it is
predominant for that of $\delta_C$. 

Equations\ (\ref{ft}) and\ (\ref{MK}) can now be
combined into a form that is convenient for the
analysis of experimental results: 

\be
\F t \equiv ft (1 + \delta_R )(1 - \delta_C ) =  
\frac{K}{2 \GV^2 (1 + \DRV )} .
\label{Ft}
\ee

\noindent Here we have defined $\F t$ as the
``corrected" $ft$-value.  From this equation, it is
evident that the $\F t$-values obtained from
$0^{+} \rightarrow 0^{+}$ transitions in different
nuclei can constitute a stringent test of CVC,
which requires them all to be equal.

To date, superallowed  $0^{+} \rightarrow 0^{+}$
transitions have been measured to $\pm 0.1\%$
precision
or better in the decays of nine nuclei ranging
from $^{10}$C to $^{54}$Co. World data on
$Q$-values, lifetimes and branching ratios -- the
results of over 100 independent measurements --
were thoroughly surveyed \cite{Ha90} in 1989 and
then updated several times since, most recently for
the WEIN98 conference \cite{WEIN98}.  The resulting
weighted averages are given in the first three columns of Table
\ref{Exres}.  Using the calculated electron-capture
probabilities \cite{Ha90} given in the next column,
we obtain the ``uncorrected" $ft$-values listed in
column 5 with partial half-lives determined from the
formula $t = t_{1/2}(1+P_{EC})/R$.

{\footnotesize
\begin{table} [t]
\begin{center}
\caption{Experimental results ($Q_{EC}$, $t_{1/2}$ and
branching
ratio, $R$), electron-capture probabilities ($P_{EC}$) and calculated corrections ($\delta_R$
and $\delta_C$)
for $0^{+} \rightarrow 0^{+}$ transitions.
\label{Exres} }
\vskip 1mm
\begin{tabular}{rcccccccc}
& \raisebox{0pt}[13pt][0pt]{$Q_{EC}$} &
\raisebox{0pt}[13pt][0pt]{$t_{1/2}$} &
\raisebox{0pt}[13pt][0pt]{$R$} &
\raisebox{0pt}[13pt][0pt]{$P_{EC}$} &
\raisebox{0pt}[13pt][0pt]{$ft$} &
\raisebox{0pt}[13pt][0pt]{$\delta_C$} &
\raisebox{0pt}[13pt][0pt]{$\delta_R$} &
\raisebox{0pt}[13pt][0pt]{$\F t$} \\
 & (keV) & (ms) & (\%) & (\%) & (s) & 
(\%) & (\%) & (s) \\[1mm]
\tableline 
\raisebox{0pt}[13pt][0pt]{$^{10}$C} &
\raisebox{0pt}[13pt][0pt]{1907.77(9)} &
\raisebox{0pt}[13pt][0pt]{19290(12)} &
\raisebox{0pt}[13pt][0pt]{1.4645(19)} &
\raisebox{0pt}[13pt][0pt]{0.296} &
\raisebox{0pt}[13pt][0pt]{3038.7(45)} &
\raisebox{0pt}[13pt][0pt]{0.16(3)} &
\raisebox{0pt}[13pt][0pt]{1.30(4)} &
\raisebox{0pt}[13pt][0pt]{3072.9(48)} \\
$^{14}$O & 2830.51(22) & 70603(18) & 99.336(10) &
0.087 & 3038.1(18) & 0.22(3) & 1.26(5) &
3069.7(26) \\
$^{26m}$Al & 4232.42(35) & 6344.9(19) & $\geq$
99.97 & 0.083 
& 3035.8(17) & 0.31(3) & 1.45(2) &
3070.0(21) \\
$^{34}$Cl & 5491.71(22) & 1525.76(88) & $\geq$
99.988 & 0.078 
& 3048.4(19) & 0.61(3) & 1.33(3) &
3070.1(24) \\
$^{38m}$K & 6044.34(12) & 923.95(64) & $\geq$
99.998 & 0.082 
& 3049.5(21) & 0.62(3) & 1.33(4) &
3071.1(27) \\
$^{42}$Sc & 6425.58(28) & 680.72(26) & 99.9941(14)
& 0.095 
& 3045.1(14) & 0.41(3) & 1.47(5) &
3077.3(23) \\
$^{46}$V & 7050.63(69) & 422.51(11) & 99.9848(13) &
0.096 & 3044.6(18) & 0.41(3) & 1.40(6) &
3074.4(27) \\
$^{50}$Mn & 7632.39(28) & 283.25(14) & 99.942(3) &
0.100 & 3043.7(16) & 0.41(3) & 1.40(7) &
3073.8(27) \\
$^{54}$Co & 8242.56(28) & 193.270(63) & 99.9955(6)
& 0.104 
& 3045.8(11) & 0.52(3) & 1.40(7) &
3072.2(27) \\
 & & & & & & \multicolumn{2}{c}{Average,
$\overline{\F t}$} & 3072.3(9)~ \\
 & & & & & & \multicolumn{2}{c}{$\chi^2/\nu$}  &
1.10 \\[1mm]
\end{tabular}
\end{center}
\end{table}
}

\begin{figure}[t]
\centerline{   
\epsfxsize=14cm
\epsfbox{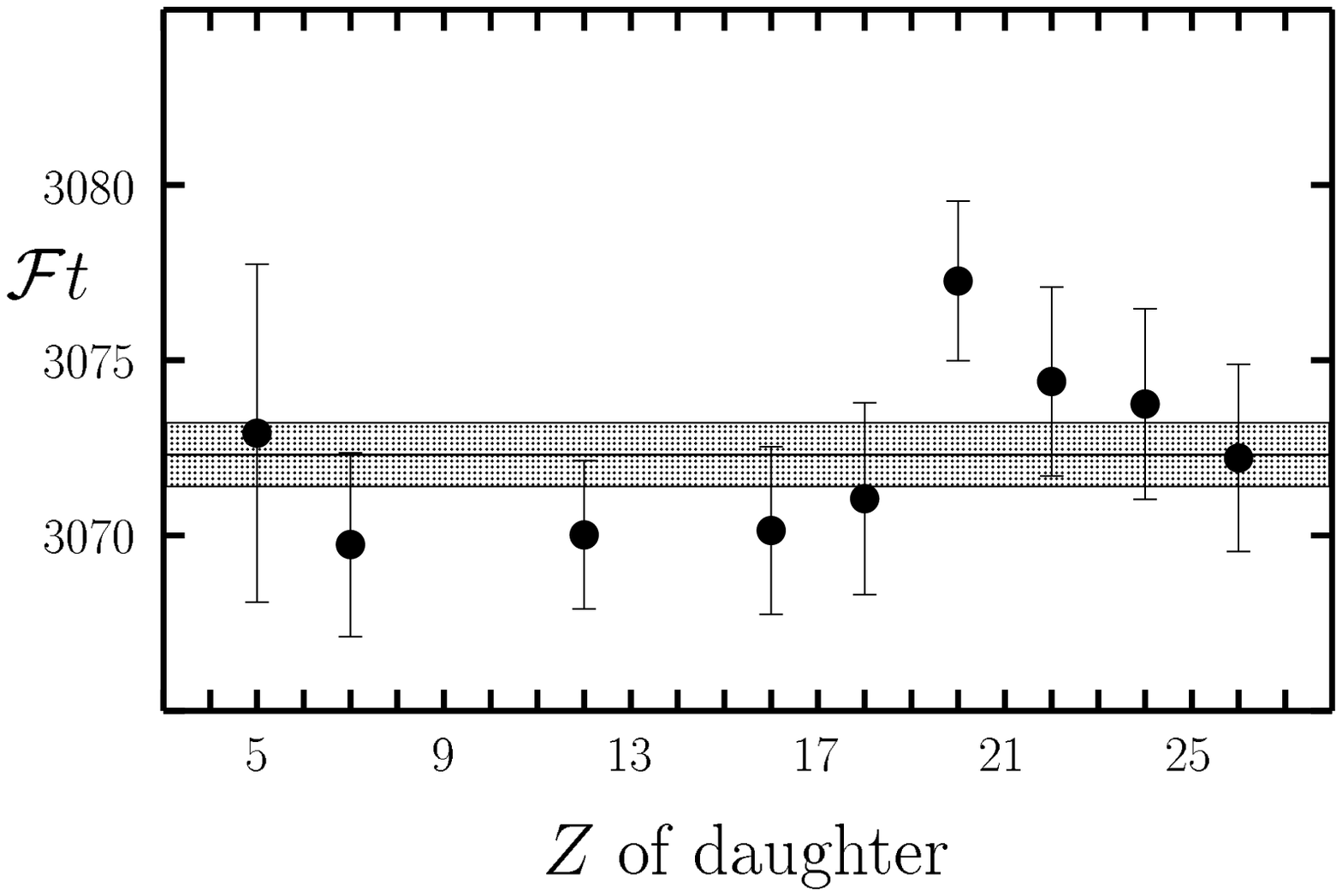}
}
\vspace{-9.5cm}
\caption{$\F t$-values for the nine precision data, and the best
least-squares one-parameter fit.  \label{fig1}}
\end{figure}

To convert these results for $ft$ into $\F t$-values, we
apply the $\delta_C$ and $\delta_R$ corrections
given in columns 6 and 7.  We describe the
$\delta_C$ calculations and their dependence on
nuclear structure in a later section of this paper; the
$\delta_C$ values used in Table \ref{Exres} were
taken from the last column of Table \ref{t:tab2}.  The
$\delta_R$ values come from our previous recent
analyses \cite{TH95,WEIN98} and arise from a
variety of primary sources
\cite{Ha90,Si67,Si87,JR87,To94}.  It is important to
appreciate that the values of $\delta_C$ and
$\delta_R$  result from more than one
independent calculation.  In
the case of $\delta_R$, the calculations are in
complete
accord with one another; for $\delta_C$, we have
used
an average of two independent calculations with
assigned uncertainties that reflect the (small) scatter
between them.  Thus, in a real sense, both
experimentally and theoretically, the $\F t$-values
given in  Table \ref{Exres} and plotted in  Fig.\
\ref{fig1} represent the totality of
current world knowledge.  The uncertainties reflect
the
experimental uncertainties and an estimate of the
{\em
relative} theoretical uncertainties in $\delta_C$. 
There is
no statistically significant evidence of inconsistencies
in
the data ($\chi^2 / \nu =
1.1$), thus verifying the expectation of CVC at the
level
of $3 \times 10^{-4}$, the fractional uncertainty
quoted
on the average $\F t$-value.

In using this average $\F t$-value to determine
$V_{ud}$ and test CKM unitarity, we must account
for additional uncertainty:
{\em viz}

\be
\overline{\F t} = 3072.3 \pm 0.9 \pm 1.1 ,
\label{Ftavg1}
\ee

\noindent where the first error is the statistical error of
the fit (as illustrated in Fig.\ \ref{fig1}),
and the second is an error related to the systematic
difference between
the two calculations of $\delta_C$ by Towner, Hardy
and
Harvey \cite{THH77,To89}
and by Ormand and Brown \cite{OB95} that we have
combined in reaching
this result.  (For a more complete discussion of
how we treat these theoretical uncertainties, see
reference \cite{Ha90}.)  We now add the two
errors linearly to obtain the value we use in
subsequent analysis:

\be
\overline{\F t} = 3072.3 \pm 2.0 .
\label{Ftavg2}
\ee

\noindent The value of $V_{ud}$ is obtained by relating
the
vector
constant,
$\GV$, determined from this $\overline{\F t}$
value, to
the weak
coupling constant from muon decay, $\GF /(\hbar
c)^3
=
(1.16639 \pm 0.00001) \times 10^{-5}$
GeV$^{-2}$, according to:

\be
V_{ud}^2 = \frac{K}{2 \GF^2 (1+\DRV )
\overline{\F
t}} .
\label{Vud2f}
\ee

\noindent With the nucleus-independent
radiative correction adopted from Sirlin \cite{Si94},
$\DRV = (2.40 \pm 0.08) \%$, we obtain the result 

\be 
\mids V_{ud} \mids \, = 0.9740 \pm 0.0005 ,
\label{Vud00}
\ee

We are now in a position to test the
unitarity of the CKM matrix by evaluating the sum of
squares of the elements in its first row.  With the
value just obtained for $V_{ud}$, combined with the
values of $V_{us}$ and $V_{ub}$ quoted by the
Particle Data Group \cite{PDG98}, the unitarity sum
becomes

\be
\mids V_{ud} \mids^2 +
\mids V_{us} \mids^2 +
\mids V_{ub} \mids^2 = 0.9968 \pm 0.0014,
\label{ist:unit}
\ee

\noindent which differs from unity at the $98\%$
confidence level.

\section*{Is non-unitarity real?}
\label{nonunit}

The result in equation\ (\ref{ist:unit}) is  a very
provocative one.  If it is taken at face
value, it indicates the need for some extension to the
electroweak Standard Model, possibly indicating
the presence of right-hand currents or of a scalar
interaction \cite{WEIN98}.  This would have
profound implications.  However, the result could
have a more trivial explanation.  It could instead
reflect some undiagnosed inadequacy in the
calculated radiative or Coulomb corrections used to
evaluate $V_{ud}$ -- or possibly a comparable
inadequacy in the evaluation of $V_{us}$.  What can
be stated with some certainty is that the experimental
results for the nine nuclei listed in Table \ref{Exres}
cannot be at fault.  Not only do they originate from a
large number of independent measurements, but also the
error bar associated with $ \, \mids V_{ud} \mids \,$ is
{\em not} predominantly experimental in origin.  In fact,
if experiment were the sole contributor, the uncertainty
would be only $\pm 0.0001$.  The largest contributions
to the $ \, \mids V_{ud} \mids \,$ error bar come from
$\DRV $ ($\pm 0.0004$) and $\delta_C$ ($\pm
0.0003$).

Thus, if we are to determine whether the minimal
Standard Model has failed, we must eliminate all
possible ``trivial" explanations for the apparent
non-unitarity.  To do so, nuclear physicists focus
on the reliability of the calculated corrections in 
$V_{ud}$.  (Others are re-evaluating $V_{us}$ -- see
reference \cite{WEIN98}.)  But, if there is a fault in the
corrections, what size effect are we seeking?  To restore
unitarity, the calculated radiative corrections ($\delta_R$
or $\DRV$) for all nine superallowed transitions would
all have to be shifted downwards by 0.3\%, or the
calculated Coulomb corrections, $\delta_C$, all shifted
upwards by 0.3\%, or some combination of the two. 
Such changes would constitute a substantial fraction of
the total values of these small quantities.  We have
recently re-examined \cite{WEIN98} the calculation of
the various correction terms to see whether such large
changes are plausible.  Our conclusion is largely
negative, based on arguments that are now briefly
presented.

The radiative correction has been conveniently divided into
terms that are nucleus-dependent, $\delta_R$, and terms that are not,
$\DRV$.  These are written

\bea
\delta_R & = & \frac{\alpha}{2 \pi} \left [ \overline{g}(E_m)
+ \delta_2 + \delta_3 + 2 C_{NS} \right ]
\nonumber \\
\DRV & = & \frac{\alpha}{2 \pi} \left [ 4 \ln (\mZ /m_p ) + \ln (m_p / \mA )
+ 2 C_{\rm Born} \right ] + \cdots ,
\label{ist:drDR}
\eea

\noindent where the ellipses represent further small terms of order
0.1\%.  In these equations, $E_m$ is the maximum electron energy
in beta decay, $\mZ$ the $Z$-boson mass, $\mA$ the $a_1$-meson mass, and
$\delta_2$ and $\delta_3$ the order $Z \alpha^2$ and $Z^2 \alpha^3$
contributions.  The electron-energy dependent function, $g(E_e,E_m)$,
was derived by Sirlin \cite{Si67}; it is here averaged over the
electron spectrum to give $\overline{g}(E_m)$.

Typical values are

\be
\delta_R \simeq 0.95 + 0.43 + 0.05 + (\alpha /\pi ) C_{NS} \% ,
\label{ist:tdr}
\ee

\noindent where $(\alpha /\pi )C_{NS}$ is of order $-0.3\%$ for
$T_z = -1$ beta emitters, $^{10}$C and $^{14}$O, and of order five
times smaller for the $T_z = 0$ emitters, ranging from
$-0.09\%$ to $+0.03\%$.  Thus for $T_z=0$ emitters
$\delta_R \simeq 1.4\%$.  If the failure to obtain unitarity in the
 CKM matrix with
$V_{ud}$ from nuclear beta decay is due to the value of $\delta_R$,
then $\delta_R$ must be reduced to 1.1\%.  This is not likely.
The leading term, 0.95\%, involves standard QED and is well
verified.  The order-$Z\alpha^2$ term, 0.43\%, while less secure
has been calculated twice \cite{Si87,JR87} independently, with
results in accord.

For the nucleus-independent term

\be
\DRV = 2.12 - 0.03 + 0.20 + 0.1\% ~~\simeq~~ 2.4\%,
\label{ist:tdrv}
\ee

\noindent of which the first term, the leading logarithm, is unambiguous.
Again, to achieve unitarity of the CKM matrix, $\DRV$ would have to be 
reduced to 2.1\%, \ie\ all terms other than the leading logarithm
summing to zero.  This also seems unlikely.

Because the leading terms in the radiative corrections
are so well founded, attention has focused more on
possible weaknesses in the Coulomb correction. 
Although smaller than the radiative correction, the
Coulomb correction is clearly sensitive to nuclear-structure
issues.  It comes about because Coulomb
and charge-dependent nuclear forces destroy isospin
symmetry between the initial and final states in
superallowed beta-decay.  The consequences are
twofold: there are different degrees of configuration
mixing in the two states, and, because their binding
energies are not identical, their radial wave functions
differ.  Thus, we accommodate both effects by
writing $\delta_C = \delta_{C1} + \delta_{C2}$.

There have been several independent calculations of
$\delta_C$.  The first followed methods developed
by Towner, Hardy and Harvey \cite{THH77} with
refinements presented in more recent publications
\cite{To89,Ha94}.  They use shell-model
calculations to determine $\delta_{C1}$, and
full-parentage expansions in terms of Woods-Saxon
radial wave functions to obtain $\delta_{C2}$.  A
second calculation, by Ormand and Brown
\cite{OB95}, also employed the shell model to
obtain $\delta_{C1}$ but derived $\delta_{C2}$
from a self-consistent Hartree-Fock calculation.  The
results of these two calculations agree remarkably
well with one another, the Towner-Hardy-Harvey
values being systematically only 0.07\% higher than
Ormand-Brown ones.  It is an average of these two
sets of values that we have used for $\delta_C$ in
our analysis as given in Table \ref{Exres}. 

Two more recent calculations provide a valuable
check that these $\delta_C$ values are not suffering
from severe systematic effects.  Sagawa, van Giai
and Suzuki \cite{SVS96} have added RPA
correlations to a Hartree-Fock calculation that
incorporates charge-symmetry and
charge-independence breaking forces in the
mean-field potential to take account of isospin
impurity in the core; the correlations, in essence,
introduce a coupling to the isovector monopole giant
resonance.  The calculation is not constrained,
however, to reproduce known separation energies as
were the two calculations already described.  Finally,
a large-basis shell-model calculation has been
mounted for the $A=10$ case by Navr\'{a}til, Barrett
and Ormand \cite{NBO97}.  Both of these two new
works have produced values of $\delta_C$ very
similar to, but actually {\em smaller} than those used
in our analysis, \ie\ worsening rather than helping the
unitarity problem.

The typical value of $\delta_C$ is of order 0.4\%.  If
the unitarity problem is to be solved by
improvements in $\delta_C$, then $\delta_C$ has to
be raised to around 0.7\%.  There is no evidence
whatsoever for such a shift from recent works.  Even
so, considerable effort has already gone into
making independent experimental checks on the
accuracy of the $\delta_C$ calculations.  These are
described in the next section.

\section*{Tests of the Coulomb correction}
\label{s:tests}

As shown in equation\ (\ref{MK}) the Coulomb
correction, $\delta_C$, modifies the square of the
nuclear matrix element: $| \MV |^2 \rightarrow | \MV
|^2 ( 1 - \delta_C )$.  Here $M_V$ is the Fermi matrix
element, the expectation value of the isospin ladder
operator, which for isospin $T=1$ states has the value
$\MV = \sqrt{2}$.  The value of $\delta_C$ clearly
depends on the detailed structure of the nuclei
involved.  In the preceding section we have written
$\delta_C$ as the sum of two components,
$\delta_{C1}$ and $\delta_{C2}$, the first reflecting the
{\em difference} between configuration mixing in the initial
and final states, while the second reflects the
differences in their radial wave functions.    Our
calculations for $\delta_{C2}$ have been documented
in a paper by Towner, Hardy and Harvey
\cite{THH77} and so will not be repeated here.  It is
sufficient to stress that constraints are placed on the
$\delta_{C2}$-calculations by insisting that the
asymptotic forms of the proton and neutron radial
functions match the known separation energies.  Our
calculations for $\delta_{C1}$ are documented here.

It is instructive to consider a simplified two-state
mixing case as it
will illustrate the issues involved.  As a specific
example, take the 
case of $^{42}$Sc decaying to $^{42}$Ca in a
superallowed Fermi transition.
In the calculation we admit two $0^{+}$ states, the
ground state and 
one excited state.  One of these states might be a
two-particle state,
$| 2p \rangle$, relative to a closed $^{40}$Ca core,
the
other might
be a four-particle two-hole state, $ | 4p \hyphen 2h
\rangle $.
The strong interaction heavily mixes the two states, so
the ground
state, $\psi_0$, and excited state, $\psi_1$, will have
wave functions

\bea
\psi_0 & = & A | 2p \rangle + B | 4p \hyphen 2h
\rangle
\nonumber \\
\psi_1 & = & B | 2p \rangle - A | 4p \hyphen 2h
\rangle ,
\label{strong_wf}
\eea

\noindent where the mixing amplitudes $A$ and $B$
will depend on the
details of the strong interaction.  The strong
interaction,
however,
is isospin invariant so the same wave functions
describe
the states
in $^{42}$Sc and the isospin mirror states in
$^{42}$Ca.
Thus the Fermi matrix element between ground states
is

\bea
\langle \psi_0 | \tau_{+} | \psi_0 \rangle & = &
A^2 \langle 2p | \tau_{+} | 2p \rangle +
B^2 \langle 4p \hyphen 2h | \tau_{+} | 4p \hyphen 2h
\rangle 
\nonumber \\
& = & \sqrt{2} ( A^2 + B^2 )
\nonumber \\
& = & \sqrt{2} ,
\label{MF0_strong}
\eea

\noindent while the Fermi matrix element between the
$^{42}$Sc ground
state and the $^{42}$Ca excited state is

\bea
\langle \psi_1 | \tau_{+} | \psi_0 \rangle & = &
A B \langle 2p | \tau_{+} | 2p \rangle -
A B \langle 4p \hyphen 2h | \tau_{+} | 4p \hyphen 2h
\rangle 
\nonumber \\
& = & \sqrt{2} ( A B - A B )
\nonumber \\
& = & 0 .
\label{MF1_strong}
\eea

\noindent This latter result is just a reminder that the
operator for 
superallowed Fermi decay, being the isospin ladder
operator, only connects
with states of the same isospin multiplet, \ie
~analogue
states.

Now, suppose we add to the Hamiltonian Coulomb
and
other charge-dependent 
forces.  These terms, no doubt, will be much weaker
than the 
strong-interaction forces, but their impact is to modify
slightly the wave
functions $\psi_0$ and $\psi_1$ and by differing
amounts in $^{42}$Sc
and $^{42}$Ca.  Thus these wavefunctions are now
written

\bea
\psi_i(^{42}Sc) & = & b_0 \psi_0 + b_1 \psi_1
\nonumber \\
\psi_{f0}(^{42}Ca) & = & a_0 \psi_0 + a_1 \psi_1
\nonumber \\
\psi_{f1}(^{42}Ca) & = & a_1 \psi_0 - a_0 \psi_1 ,
\label{mixed_wf}
\eea

\noindent with $a_0$ and $b_0$ both being close to
unity.  The Fermi 
matrix element between ground states becomes

\bea
\langle \psi_{f0}(^{42}Ca) | \tau_{+} |
\psi_i(^{42}Sc)
\rangle & = &
a_0 b_0 \langle \psi_0 | \tau_{+} | \psi_0 \rangle +
a_1 b_1 \langle \psi_1 | \tau_{+} | \psi_1 \rangle 
\nonumber \\
& = & \sqrt{2} ( a_0 b_0 + a_1 b_1 ) ,
\nonumber \\
| \langle \psi_{f0}(^{42}Ca) | \tau_{+} |
\psi_i(^{42}Sc) \rangle |^2 
& \simeq & 2 \left ( 1 - (a_1 - b_1 )^2 \right ) ,
\label{MF0_mixed}
\eea

\noindent using $a_0^2 + a_1^2 = 1$,
$b_0^2 + b_1^2 = 1$.  Further the matrix element
between $^{42}$Sc
ground state and the $^{42}$Ca excited state is

\bea
\langle \psi_{f1}(^{42}Ca) | \tau_{+} |
\psi_i(^{42}Sc)
\rangle & = &
a_1 b_0 \langle \psi_0 | \tau_{+} | \psi_0 \rangle -
a_0 b_1 \langle \psi_1 | \tau_{+} | \psi_1 \rangle 
\nonumber \\
& = & \sqrt{2} ( a_1 b_0 - a_0 b_1 ) ,
\nonumber \\
| \langle \psi_{f1}(^{42}Ca) | \tau_{+} |
\psi_i(^{42}Sc) \rangle |^2 
& \simeq & 2  (a_1 - b_1 )^2 ,
\label{MF1_mixed}
\eea

\noindent and is no longer zero.  If we write the
ground
state to ground 
state matrix element squared as: $| \MV^0 |^2 = 2 ( 1
-
\delta_{C1}^0 )$,
and the ground state to excited state matrix element
squared as:
$| \MV^1 |^2 = 2 \delta_{C1}^1$, then for two-state
mixing the
corrections are equal:

\be
\delta_{C1}^0 = \delta_{C1}^1 = (a_1 - b_1 )^2 .
\label{2_state}
\ee

\noindent This is a specific result for two-state
mixing,
in general they
would not be equal.  Further the correction
$\delta_{C1}$ depends on
the {\it difference} in the degree of isospin mixing in
$^{42}$Sc relative
to $^{42}$Ca.  This is clearly evident if we use
perturbation theory
to estimate the small amplitudes $a_1$ and $b_1$. 
Let
$V_C (^{42}Ca )$ and
$V_C (^{42}Sc )$ be the Coulomb and other
charge-dependent forces
operative in $^{42}$Ca and $^{42}$Sc respectively,
then

\be
\delta_{C1}^0 = \delta_{C1}^1 = 
\frac{\left [ \langle \psi_1 | V_C(^{42}Ca) | \psi_0
\rangle -
\langle \psi_1 | V_C(^{42}Sc) | \psi_0 \rangle \right
]^2
}
{(E_1 - E_0)^2} ,
\label{2_state_p}
\ee

\noindent where $E_1 - E_0$ is the energy separation
between the excited-
and ground- $0^{+}$ states from the
charge-independent Hamiltonian.
Thus $\delta_{C1}$ is inversely proportional to the
square of this
energy difference.

In a shell-model calculation it is notoriously difficult
to
obtain correctly
the experimental $E_1 - E_0$ energy separation. 
This
is because the
$| 4p \hyphen 2h \rangle$ is a deformed state, while
the
$| 2p \rangle$ is a spherical state and these two
distinct
aspects are
difficult to realise in a truncated calculation in a
spherical basis.
Thus in the calculations we are about to describe, the
values of
$\delta_{C1}^0$ and
$\delta_{C1}^1$ obtained are both corrected using

\be
\delta_{C1} = \delta_{C1}^{{\rm calc}} \times
\frac{ ( E_1 - E_0 )^2_{{\rm calc}}}
{ ( E_1 - E_0 )^2_{{\rm expt}}} .
\label{corr}
\ee

\noindent In this way, we believe we are adjusting the
$\delta_{C1}$
value approximately for the imperfections of the
underlying 
strong-interaction Hamiltonian and the necessity of
using model-space
truncations.

Here we discuss calculations for $^{38m}$K,
$^{42}$Sc, $^{46}$V,
$^{50}$Mn and $^{54}$Co, which were recomputed
recently to compare with the
excited state non-analogue Fermi transitions
measured
by
Hagberg \etal \cite{Ha94}.
In each case we use the largest model space
practicable
in a
proton-neutron (pn) basis.  For $^{38m}$K and
$^{42}$Sc
this involved orbitals in both the $(s,d)$ and $(p,f)$
shells and the effective interaction constructed by
Warburton, Becker, Millener and Brown (WBMB)
\cite{WBMB}
was used for the underlying strong interaction.  For
$^{46}$V, $^{50}$Mn and $^{54}$Co, the orbitals
span the
$(p,f)$ shell and the strong interaction was taken to be
FPMI3 from Richter \etal \cite{FPMI3}.
The single-particle energies were fixed from
experimental
values at the closed-shell-plus-one configuration.

The Coulomb and charge-dependent interaction terms
to be added were
constrained so that the ground-state masses of the
isotriplet
of states, $T_z = -1,0,+1$, were fitted.  The isobaric
multiplet mass
equation (IMME) writes these masses as

\be
M(T_z) = a + b T_z + c T_z^2 ,
\label{IMME}
\ee

\noindent where the coefficient $a$ represents the
result
from a
charge-independent Hamiltonian, and $b$ and $c$ are
charge-dependent
corrections.  Specifically

\bea
b & = & \left ( M(T_z = +1) - M(T_z = -1) \right ) /2
\nonumber \\
c & = & \left ( M(T_z = +1) + M(T_z = -1) - 2 M(T_z
=
0) \right ) /2 ,
\label{bc_coef}
\eea

\noindent so $b$ is governed by the difference
between
pp and nn
forces, and $c$ by the difference between the pn
and
nn forces.
Our strategy was to multiply the two-body Coulomb
matrix elements by a
constant factor so that the $b$-coefficient of the
IMME
is reproduced,
and likewise the pn matrix elements are multiplied
by a constant
factor to reproduce the $c$-coefficient.  This strategy
of
adjusting
the strength of the Coulomb and charge-dependent
nuclear forces to
reproduce the IMME equation was pioneered by
Ormand and Brown \cite{OB85}.

\begin{table*}[t!]
\caption{Experimental branching ratios, $R$, for
non-analogue Fermi
transitions, and values of $\delta_{C1}$ from
experiment and theory.}
\label{t:tab1}
\begin{tabular}{rcccc}
& \multicolumn{2}{c}{Expt\tablenote{From
Hagberg
\etal \cite{Ha94}}}
& \multicolumn{2}{c}{Theory}  \\

Nuclide & R(ppm) & $\delta_{C1}^1 (\% )$ &
$\delta_{C1}^1 (\% )$ &
$\delta_{C1}^0 (\% )$ \\[1mm]
\tableline
$^{38m}$K & $<19$ & $<0.28$ & 0.096 & 0.100 \\
$^{42}$Sc & 59(14)\tablenote{Daehnick \etal
\cite{Da85} averaged
with earlier results \cite{Sa80,De78,In77}} &
0.040(9)
& 0.041 & 0.049 \\
$^{46}$V & 39(4) & 0.053(5) & 0.046 & 0.087 \\
$^{50}$Mn & $<3$ & $<0.016$ & 0.051 & 0.068 \\
$^{54}$Co & 45(6) & 0.035(5) & 0.037 & 0.045
\\[1mm]
\end{tabular}
\end{table*}

The results of these calculations for $\delta_{C1}^0$
and $\delta_{C1}^1$ for nuclei $^{38m}$K,
$^{42}$Sc, $^{46}$V,
$^{50}$Mn and $^{54}$Co are given in Table
\ref{t:tab1}.
Only the $\delta_{C1}^0$ value is needed for the
analysis of the superallowed
Fermi data.  However, the companion
$\delta_{C1}^1$ value can be subjected to an
experimental test.
If the Fermi transition to an excited $0^{+}$,
non-analogue,
state can be determined then the measured branching
ratio, $R$,
is proportional to
$\delta_{C1}^1$.  These branching ratios are very
small, parts per
million (ppm), so their measurement requires a
dedicated effort, and
the work of the Chalk River group in this regard is
described
in Hagberg \etal \cite{Ha94}.  If $t_0$ is the partial
half-life
for the ground-state decay and $t_1$ the the partial
half-life to the
excited $0^{+}$ state, then

\be
R = \frac{t_0}{t_1} =
\frac{f_1}{f_0} \frac{f_0 t_0}{f_1 t_1} =
\frac{f_1}{f_0} \frac{2 \delta_{C1}^1}{2(1-
\delta_{C1}^0)} \simeq
\frac{f_1}{f_0} \delta_{C1}^1 ,
\label{ratio_R}
\ee

\noindent where $f_0$ and $f_1$ are phase space
integrals for the
ground state and excited state respectively.  Table
\ref{t:tab1}
lists the experimental value of $R$ and
$\delta_{C1}^1$.  The
comparison between theory and experiment is
exceptional in all cases 
except $^{50}$Mn and so provides a lot of
confidence
that the
companion $\delta_{C1}^0$ values used in the
superallowed
Fermi data analysis are reasonable.  The $^{50}$Mn
case is interesting
in that the experiment was unable to locate a branch
to
an excited $0^{+}$
state in the expected energy region and so deduced
that
any such 
branching ratio would be less than 3 ppm.

\begin{table*}[t!]
\caption{Calculated Coulomb correction, $\delta_C$
in
percent units.}
\label{t:tab2}
\begin{tabular}{rccccc}
&
\multicolumn{2}{c}{Towner-Hardy\tablenote{Refs.
\cite{THH77,To89}}}
&
\multicolumn{2}{c}{Ormand-Brown\tablenote{Ref.
\cite{OB95}}}  &               
Adopted Value\tablenote{Average of Towner-Hardy
and
Ormand-Brown values; assigned
uncertainties
reflect the {\it relative} scatter between these
calculations.} \\
Nuclide &  $\delta_{C1}^0 (\% )$ &
$\delta_{C2} (\% )$ &
$\delta_{C1}^0 (\% )$ &
$\delta_{C2} (\% )$ &
$\delta_C = \delta_{C1} + \delta_{C2} (\%)$
\\[1mm]
\tableline
$^{10}$C & 0.006 & 0.17 & 0.04 & 0.11 & 0.16(3) \\
$^{14}$O & 0.004 & 0.28 & 0.01 & 0.14 & 0.22(3)
\\
$^{26m}$Al & 0.057 & 0.27 & 0.01 & 0.29 &
0.31(3)
\\
$^{34}$Cl & 0.024 & 0.62 & 0.06 & 0.51 & 0.61(3)
\\
$^{38m}$K & 0.100 & 0.54 & 0.11 & 0.48 & 0.62(3)
\\
$^{42}$Sc & 0.049 & 0.35 & 0.11 & 0.31 & 0.41(3)
\\
$^{46}$V & 0.087 & 0.36 & 0.09 & 0.29 & 0.41(3)
\\
$^{50}$Mn & 0.068 & 0.40 & 0.02 & 0.33 & 0.41(3)
\\
$^{54}$Co & 0.045 & 0.56 & 0.04 & 0.40 & 0.52(3)
\\[1mm]
\end{tabular}
\end{table*}

The $\delta_{C1}^0$ values for lighter superallowed
Fermi emitters,
$^{10}$C, $^{14}$O, $^{26m}$Al and $^{34}$Cl
were obtained in a similar
manner, as described in Towner \cite{To89}.  The 
$\delta_{C1}$ = $\delta_{C1}^0$ values are quite
small, of the order
of $0.1 \%$, but seem under reasonable control.  The
larger $\delta_{C2}$
values associated with radial overlap integrals,
potentially can have more
uncertainty.  We have found that our results based on
using
Saxon-Woods radial functions lead to systematically
larger
$\delta_{C2}$
values than the calculations of Ormand and Brown
\cite{OB95} with
Hartree-Fock functions.  We therefore allow for this
systematic difference in our data analysis by using
average values and introducing a systematic
uncertainty to $\overline{\F t}$ (see equation\
(\ref{Ftavg1})).  In Table
\ref{t:tab2}
we give our $\delta_{C1}$ and $\delta_{C2}$ values,
together
with those of Ormand and Brown, and our adopted
final
numbers,
$\delta_C = \delta_{C1} + \delta_{C2}$.  These are the
values that appear in Table \ref{Exres} and are used
to analyze the world data for superallowed $0^{+}
\rightarrow 0^{+}$ decays.

\section*{Future Directions}
\label{future}

     With the experimental evidence so far completely in
support of the calculated values for $\delta_C$, the
current world data on superallowed $0^{+}\rightarrow
0^{+}$ beta decay are tantalizingly close to a result in
definitive disagreement with CKM unitarity.  Naively
one might expect such a situation would prompt an
urgent new round of experiments with the goal of
further reducing the quoted uncertainty in $ \, \mids
V_{ud} \mids \,$ but, unfortunately, the next step
cannot be so straightforward.  As we have already
noted, the error bar associated with $ \, \mids V_{ud}
\mids \,$ in equation\ (\ref{Vud00}) is now dominated
by uncertainties in the calculated correction terms.  Any
improvements in precision made within the existing
body of experimental data will be effectively lost once
the results are applied to the unitarity test so long as
there are no improvements in the calculations.

     Clearly of highest priority in future must be to
increase the precision of the correction terms,
particularly $\DRV $, which is the largest contributor to
the uncertainty of $ \, \mids V_{ud} \mids \,$.  Its
importance is manifest not only in the unitarity test
based on superallowed $0^{+}\rightarrow 0^{+}$ beta
decay but also in any tests based on neutron or pion
decays.  To date, the experimental data on these
non-nuclear decays are considerably less
precise
\cite{WEIN98} than those on the $0^{+}\rightarrow
0^{+}$ transitions but, year by year, improvements are
being made in the neutron-decay measurements largely
motivated by the prospect of a unitarity test unfettered
with the structure-dependent Coulomb correction,
$\delta_C$, which vanishes for the neutron.  In spite of
this simplification in the neutron decay, however, any
potential advantage to the unitarity test is lost unless
$\DRV $ can be calculated with greater precision, since
{\em all} determinations of $ \, \mids V_{ud} \mids \,$
depend directly on $\DRV $.

     Until such time as the $\DRV $ calculation is refined
and the neutron decay measurements rival the precision
of the nuclear results, the best hope for improvements
to the unitarity test lies in increasing our confidence in
the calculated values of $\delta_C$.  Though a
reasonable estimate of the uncertainty in $\delta_C$ has
been incorporated into the derivation of $ \, \mids
V_{ud} \mids \,$ (see Tables \ref{Exres} and
\ref{t:tab2}), there is no doubt that the dominant role of
nuclear structure in the calculation of $\delta_C$ leaves
some people questioning whether the real uncertainty is
larger than the one actually quoted.  Under the
circumstances, any experiment that can probe the
veracity of the $\delta_C$ calculations will make a
valuable contribution to the whole problem.  There are
at least three different approaches that can be taken in
devising such experiments.

     First, improvements can be sought in results for the
nine superallowed transitions whose $ft$-values are
already known to within a fraction of a percent.  This would
not be a fruitless endeavour.  It
is certainly true that, given the large quantity of careful
measurements now contributing to the content of Table
\ref{Exres}, there is little chance that the central value
of $\overline{\F t}$ will be changed significantly by a
few more.  And, it is also true that improvements in the
experimental uncertainties will not be directly reflected
in a reduced uncertainty for $ \, \mids V_{ud} \mids \,$
unless the $\DRV $ correction has been improved
too.  But, the test of CVC can be made more demanding
as the experimental precision is increased and, to the
extent that the $\F t$-values continue to agree with one
another, this would demonstrate at the same time the
reliability of the $\delta_C$ calculations, which
compensate for the transition-to-transition variations
evident in the uncorrected $ft$-values.  Of course, it is
only the {\em relative} values of $\delta_C$ that can be
tested by this method, but it would be a pathological
fault indeed that could calculate in detail the required
variations in
$\delta_C$ while failing to obtain their {\em
absolute}
values to comparable precision.

\begin{figure}[t]
\vspace{-3.5cm}
\centerline{   
\epsfxsize=14cm
\epsfbox{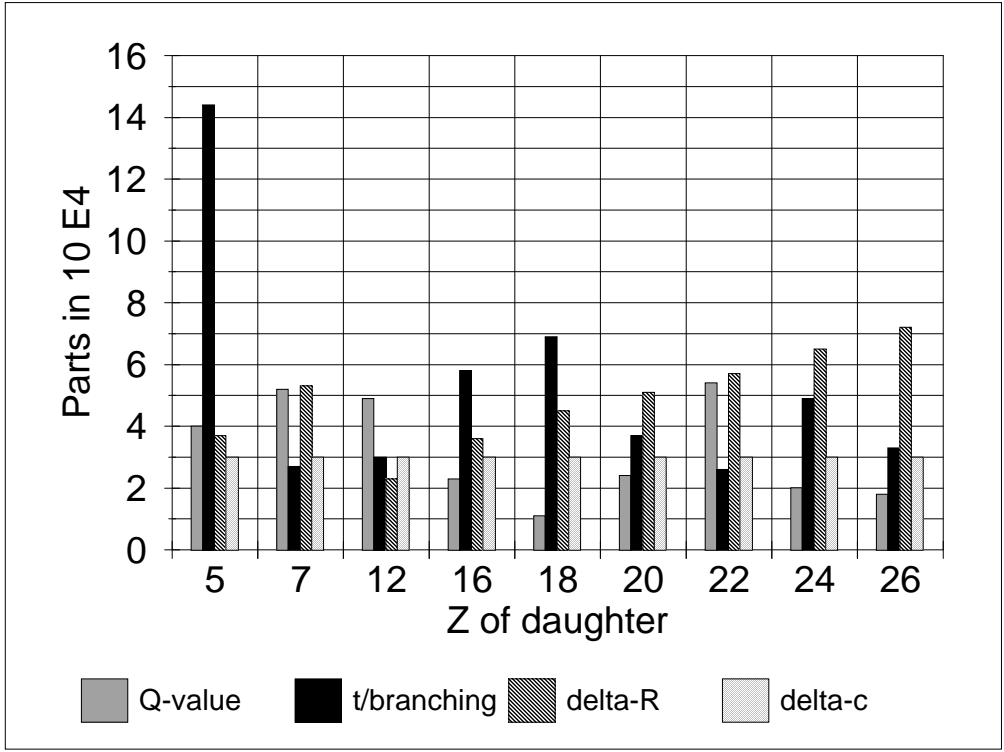}
}
\vspace{-3.8cm}
\caption{Contributions from experiment and theory to
the overall $\F t$-value uncertainty for each
superallowed transition listed in Table \ref{Exres}.  
\label{fig2}}
\end{figure}

The various experimental and theoretical contributions to
the $\F t$-value uncertainties are shown in Fig.\
\ref{fig2} for the nine superallowed transitions whose
$ft$ values are known to within a fraction of a percent. 
If we accept that it is valuable for experiment to be at
least a factor of two more precise than the calculations
for $\delta_R$ and $\delta_C$ ({\em relative}
uncertainties only), then an examination of Fig.\
\ref{fig2} shows that the Q-values for
$^{10}$C, $^{14}$O, $^{26m}$Al and
$^{46}$V, the half-lives of $^{10}$C,
$^{34}$Cl and $^{38m}$K, and the
branching ratio for $^{10}$C can all bear
improvement.  Such improvements will soon
be feasible.  The Q-values will reach 
the required level (and more) as mass
measurements with new on-line Penning traps
become possible; half-lives will likely yield to
measurements with higher statistics as 
high-intensity beams of separated isotopes are
developed for radioactive-beam
facilities; and, finally, an improved 
branching-ratio measurement on $^{10}$C has
already
been made with Gammasphere and simply
awaits analysis \cite{FR98}. 

\begin{figure}[t]
\vspace{-1.0cm}
\centerline{   
\epsfxsize=14cm
\epsfbox{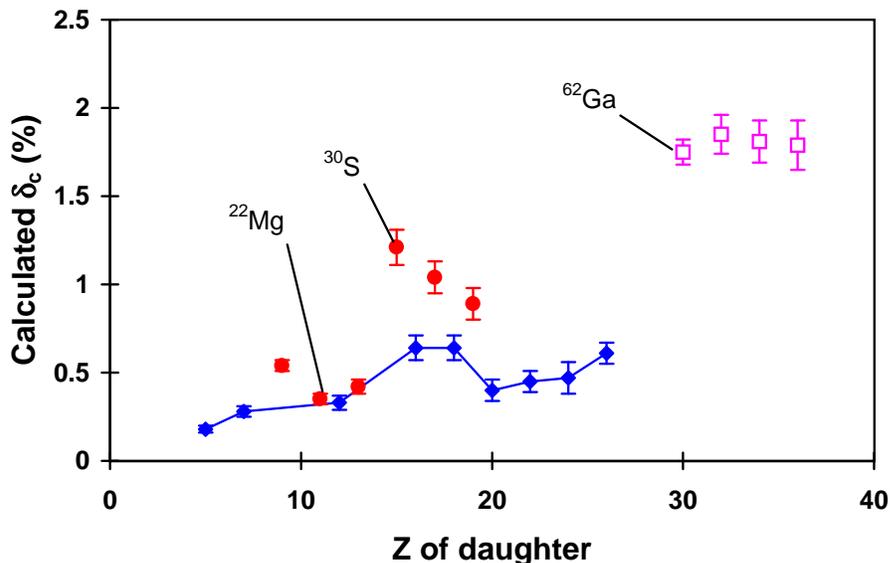}
}
\vspace{-1.5cm}
\caption{Calculated $\delta_C$ values plotted as a
function of the Z of the daughter nucleus.  The solid
diamonds joined by the line represent the values for
the nine well known superallowed emitters listed in
Table \ref{Exres}.  The circles are for the $T_z =
-1$ emitters between $^{18}$Ne and $^{38}$Ca; while
the
open squares are for the $T_z = 0$ cases from
$^{62}$Ga to $^{74}$Rb. 
\label{fig3}}
\end{figure}

Another experimental approach to testing $\delta_C$ is
offered by the possibility of increasing the number of
superallowed emitters 
accessible to precision studies.  The greatest
attention recently has been paid to the $T_z =
0$ (odd-odd) emitters with $A \geq 62$, since these
nuclei are expected to be produced at new
radioactive-beam facilities, and their
calculated Coulomb corrections, $\delta_C$,
are predicted to be large
\cite{OB95,SVS96,JH92}, as is illustrated in Fig.\
\ref{fig3}.  In principle, then, they could provide a
valuable test of the accuracy of
$\delta_C$ calculations.  It is likely,
though, that these heavy emitters will not provide
$ft$-values with sufficient precision to be useful
directly in extracting competitive $\F t$-values in the
near future. 
All of the well
known emitters listed in Table \ref{Exres}, with the
exception of $^{10}$C, have the special advantage that
the superallowed branch from each is by far the
dominant transition in its decay ($>$ 99\%).  This
means that the branching ratio for the 
superallowed transitions can be
determined to high precision from relatively
imprecise measurements of the other weak
transitions, which can simply be
subtracted from 100\%.  In contrast, the decays of the
heavier  $T_z = 0$ emitters -- nuclides such as
$^{62}$Ga, $^{66}$As, $^{70}$Br and $^{74}$Rb -- will be of
considerably higher energy and
each will therefore involve several allowed
transitions of significant intensity in addition
to the superallowed transition.  Branching-ratio 
measurements will thus be very
demanding, particularly with the limited
radioactive-beam intensities likely to be available
initially for
these rather exotic nuclei.  Lifetime
measurements will be similarly constrained by
statistics.  As to the $Q_{EC}$ values, even Penning
traps will be hard pressed to
produce the required precision of a few parts in 10$^9$
for the masses of these short-lived ($t_{1/2} \leq 100$
ms) nuclides.

More immediately achievable among these heavier
superallowed emitters are measurements of the type
described in the preceding section, in which
non-analogue Fermi transitions were observed
\cite{Ha94} and compared with model calculations as
a test of the techniques used in calculating $\delta_C$. 
Such measurements would yield important information
on Coulomb mixing and they would be an important
prerequisite for any serious attempt to obtain $\F
t$-values in this region of nuclei.  Another important
prerequisite would be the experimental determination
of the coefficients of the IMME for all relevant $0^+$
$T = 1$ states.  These demanding experiments are
interesting in their own right but, as preliminaries to
future $\F t$-value determinations, they are especially
important because model calculations in this region of
rapid shape changes are likely to be far less reliable
than they are in the $(s,d)$ and $(p,f)$ shells.  Without
the constraints provided by these experiments -- and
possibly even with them -- tests of $\delta_C$ from the
measured $\F t$-values could end up reflecting more
about the limitations of the nuclear models used than
about the underlying physics of the weak interaction. 
As a first probe of these issues, an investigation of
non-analogue Fermi transitions from
$^{62}$Ga is already underway \cite{Hy98}.  More
will undoubtedly follow.

In the near future, the most promising experimental
approach that can actually increase the number of
precisely measured $\F t$-values is to study the $T_z =
-1$ superallowed emitters with $18 \leq A \leq 38$. 
There is good reason to explore them.  For example, as
shown in  Fig.\ \ref{fig3}, the calculated value
\cite{THH77} of $\delta_C$ for $^{30}$S decay,
though smaller than the $\delta_C$'s expected
for the heavier nuclei, is actually 1.2\% -- 
about a factor of two larger than for any other
case currently known -- while $^{22}$Mg
has a very low value of 0.35\%.  If the $ft$-values for
these two nuclei can be determined to a precision of a
few tenths of a percent or better, and the large
predicted difference is confirmed, then it will do much
to increase our confidence in the calculated Coulomb
corrections.  This would be especially convincing
since the calculation involves the same model space as
was used for the presently known cases.   To be sure,
these decays will provide an experimental challenge,
particularly in the measurement of their branching
ratios, but the required precision should be achievable
with isotope-separated beams that are currently
available.  In fact, such experiments are also in their
early stages at the Texas A\&M cyclotron \cite{Ha98}. 

\section*{Conclusions}
\label{conc}

The current world data on superallowed $0^+
\rightarrow 0^+$ beta decays lead to a
self-consistent
set of $\F t$-values that agree with CVC but
differ
provocatively, though not yet definitively, from
the
expectation of CKM unitarity.  There are no
evident
defects in the calculated radiative and Coulomb
corrections that could remove the problem, but
suspicion continues to fall on the calculations of
Coulomb mixing, which depend sensitively on the
details of nuclear structure.  If any progress is to be
made in firmly establishing (or eliminating) the
discrepancy with unitarity, additional experiments are
required that focus on this issue.  We have indicated
what some relevant nuclear experiments might be, and
have particularly emphasized that experiments to
measure the $ft$-values of heavy $T_z = 0$ odd-odd
superallowed emitters with $A \geq 62$, which have
been proposed for new radioactive-beam facilities, are
very difficult and should be preceded by measurements
in the same mass region of non-analogue Fermi decays
and IMME coefficients.

On the theoretical side, the most important
requirement for all tests of CKM universality that
depend upon $V_{ud}$ is an improved determination
of the nucleus-independent radiative correction,
$\DRV $.  It is not only the results from nuclear
superallowed decays that must be subjected to this
correction term, but also the results from the neutron
and pion decays if they also are to be used to extract
$V_{ud}$.  Though the latter decays are currently
known with less precision than the nuclear decays, one
can reasonably expect them to improve significantly
over the next decade.  Thus, it is of highest priority to
reduce the uncertainty currently attached to the
calculation of $\DRV $.  That having been stated, it
must be noted that nuclear decays will also require
more reliable $\delta_C$ calculations to remain
competitive, especially if the uncertainties are reduced
on $\DRV $.  In any case, improved model
calculations of nuclear structure and Coulomb mixing
in nuclei with $N \simeq Z$ and $A \geq 62$ are an
important requirement for the future if $ft$-value
measurements are to be attempted in this region.

The work of JCH was supported by
the U.S. Department of Energy under Grant
number DE-FG05-93ER40773 and by the Robert
A. Welch Foundation.

\end{document}